\documentclass[12pt,a4]{article}
\usepackage{amsfonts}
\usepackage[dvips]{epsfig}
\usepackage[dvips]{graphics}

\newcommand{\text}{\rm}

\makeatletter
\newcount\@hour\newcount\@minute\chardef\@x10\chardef\@xv60
\def\tcitime{
\def\@time{%
  \@minute\time\@hour\@minute\divide\@hour\@xv
  \ifnum\@hour<\@x 0\fi\the\@hour:%
  \multiply\@hour\@xv\advance\@minute-\@hour
  \ifnum\@minute<\@x 0\fi\the\@minute
  }}%
\def\QCTOpt[#1]#2{%
  \def\QCTOptB{#1}
  \def\QCTOptA{#2}
}
\def\QCTNOpt#1{%
  \def\QCTOptA{#1}
  \let\QCTOptB\empty
}
\def\Qct{%
  \@ifnextchar[{%
    \QCTOpt}{\QCTNOpt}
}
\def\QCBOpt[#1]#2{%
  \def\QCBOptB{#1}
  \def\QCBOptA{#2}
}
\def\QCBNOpt#1{%
  \def\QCBOptA{#1}
  \let\QCBOptB\empty
}
\def\Qcb{%
  \@ifnextchar[{%
    \QCBOpt}{\QCBNOpt}
}
\def\PrepCapArgs{%
  \ifx\QCBOptA\empty
    \ifx\QCTOptA\empty
      {}%
    \else
      \ifx\QCTOptB\empty
        {\QCTOptA}%
      \else
        [\QCTOptB]{\QCTOptA}%
      \fi
    \fi
  \else
    \ifx\QCBOptA\empty
      {}%
    \else
      \ifx\QCBOptB\empty
        {\QCBOptA}%
      \else
        [\QCBOptB]{\QCBOptA}%
      \fi
    \fi
  \fi
}
\newcount\GRAPHICSTYPE
\GRAPHICSTYPE=\z@
\def\GRAPHICSPS#1{%
 \ifcase\GRAPHICSTYPE
   \special{ps: #1}%
 \or
   \special{language "PS", include "#1"}%
 \fi
}%
\def\graffile#1#2#3#4{%
    \leavevmode
    \raise -#4 \BOXTHEFRAME{%
        \hbox to #2{\raise #3\hbox{\null #1}}}%
}%
\def\draftbox#1#2#3#4{%
 \leavevmode\raise -#4 \hbox{%
  \frame{\rlap{\protect\tiny #1}\hbox to #2%
   {\vrule height#3 width\z@ depth\z@\hfil}%
  }%
 }%
}%
\newcount\draft
\draft=\z@

\newif\ifwasdraft
\wasdraftfalse

\def\GRAPHIC#1#2#3#4#5{%
 \ifnum\draft=\@ne\draftbox{#2}{#3}{#4}{#5}%
  \else\graffile{#1}{#3}{#4}{#5}%
  \fi
 }%
\def\addtoLaTeXparams#1{%
    \edef\LaTeXparams{\LaTeXparams #1}}%
\newif\ifBoxFrame \BoxFramefalse
\newif\ifOverFrame \OverFramefalse
\newif\ifUnderFrame \UnderFramefalse

\def\BOXTHEFRAME#1{%
   \hbox{%
      \ifBoxFrame
         \frame{#1}%
      \else
         {#1}%
      \fi
   }%
}

\def\doFRAMEparams#1{\BoxFramefalse\OverFramefalse\UnderFramefalse\readFRAMEparams#1\end}%
\def\readFRAMEparams#1{%
 \ifx#1\end%
  \let\next=\relax
  \else
  \ifx#1i\dispkind=\z@\fi
  \ifx#1d\dispkind=\@ne\fi
  \ifx#1f\dispkind=\tw@\fi
  \ifx#1t\addtoLaTeXparams{t}\fi
  \ifx#1b\addtoLaTeXparams{b}\fi
  \ifx#1p\addtoLaTeXparams{p}\fi
  \ifx#1h\addtoLaTeXparams{h}\fi
  \ifx#1X\BoxFrametrue\fi
  \ifx#1O\OverFrametrue\fi
  \ifx#1U\UnderFrametrue\fi
  \ifx#1w
    \ifnum\draft=1\wasdrafttrue\else\wasdraftfalse\fi
    \draft=\@ne
  \fi
  \let\next=\readFRAMEparams
  \fi
 \next
 }%
\def\IFRAME#1#2#3#4#5#6{%
      \bgroup
      \let\QCTOptA\empty
      \let\QCTOptB\empty
      \let\QCBOptA\empty
      \let\QCBOptB\empty
      #6%
      \parindent=0pt%
      \leftskip=0pt
      \rightskip=0pt
      \setbox0 = \hbox{\QCBOptA}%
      \@tempdima = #1\relax
      \ifOverFrame
          \typeout{This is not implemented yet}%
          \show\HELP
      \else
         \ifdim\wd0>\@tempdima
            \advance\@tempdima by \@tempdima
            \ifdim\wd0 >\@tempdima
               \textwidth=\@tempdima
               \setbox1 =\vbox{%
                  \noindent\hbox to \@tempdima{\hfill\GRAPHIC{#5}{#4}{#1}{#2}{#3}\hfill}\\%
                  \noindent\hbox to \@tempdima{\parbox[b]{\@tempdima}{\QCBOptA}}%
               }%
               \wd1=\@tempdima
            \else
               \textwidth=\wd0
               \setbox1 =\vbox{%
                 \noindent\hbox to \wd0{\hfill\GRAPHIC{#5}{#4}{#1}{#2}{#3}\hfill}\\%
                 \noindent\hbox{\QCBOptA}%
               }%
               \wd1=\wd0
            \fi
         \else
            \ifdim\wd0>0pt
              \hsize=\@tempdima
              \setbox1 =\vbox{%
                \unskip\GRAPHIC{#5}{#4}{#1}{#2}{0pt}%
                \break
                \unskip\hbox to \@tempdima{\hfill \QCBOptA\hfill}%
              }%
              \wd1=\@tempdima
           \else
              \hsize=\@tempdima
              \setbox1 =\vbox{%
                \unskip\GRAPHIC{#5}{#4}{#1}{#2}{0pt}%
              }%
              \wd1=\@tempdima
           \fi
         \fi
         \@tempdimb=\ht1
         \advance\@tempdimb by \dp1
         \advance\@tempdimb by -#2%
         \advance\@tempdimb by #3%
         \leavevmode
         \raise -\@tempdimb \hbox{\box1}%
      \fi
      \egroup%
}%
\def\DFRAME#1#2#3#4#5{%
 \begin{center}
     \let\QCTOptA\empty
     \let\QCTOptB\empty
     \let\QCBOptA\empty
     \let\QCBOptB\empty
     \ifOverFrame 
        #5\QCTOptA\par
     \fi
     \GRAPHIC{#4}{#3}{#1}{#2}{\z@}
     \ifUnderFrame 
        \par #5\QCBOptA
     \fi
 \end{center}%
 }%
\def\FFRAME#1#2#3#4#5#6#7{%
 \begin{figure}[#1]%
  \let\QCTOptA\empty
  \let\QCTOptB\empty
  \let\QCBOptA\empty
  \let\QCBOptB\empty
  \ifOverFrame
    #4
    \ifx\QCTOptA\empty
    \else
      \ifx\QCTOptB\empty
        \caption{\QCTOptA}%
      \else
        \caption[\QCTOptB]{\QCTOptA}%
      \fi
    \fi
    \ifUnderFrame\else
      \label{#5}%
    \fi
  \else
    \UnderFrametrue%
  \fi
  \begin{center}\GRAPHIC{#7}{#6}{#2}{#3}{\z@}\end{center}%
  \ifUnderFrame
    #4
    \ifx\QCBOptA\empty
      \caption{}%
    \else
      \ifx\QCBOptB\empty
        \caption{\QCBOptA}%
      \else
        \caption[\QCBOptB]{\QCBOptA}%
      \fi
    \fi
    \label{#5}%
  \fi
  \end{figure}%
 }%
\newcount\dispkind%
\def\FRAME#1#2#3#4#5#6#7#8{%
 \ifnum\draft=\@ne
   \wasdrafttrue
 \else
   \wasdraftfalse%
 \fi
 \def\LaTeXparams{}%
 \dispkind=\z@
 \def\LaTeXparams{}%
 \doFRAMEparams{#1}%
 \ifnum\dispkind=\z@\IFRAME{#2}{#3}{#4}{#7}{#8}{#5}\else
  \ifnum\dispkind=\@ne\DFRAME{#2}{#3}{#7}{#8}{#5}\else
   \ifnum\dispkind=\tw@
    \edef\@tempa{\noexpand\FFRAME{\LaTeXparams}}%
    \@tempa{#2}{#3}{#5}{#6}{#7}{#8}%
    \fi
   \fi
  \fi
  \ifwasdraft\draft=1\else\draft=0\fi{}%
 }%
\def\TEXUX#1{"texux"}

\long\def\QQQ#1#2{%
     \long\expandafter\def\csname#1\endcsname{#2}}%
\@ifundefined{QTP}{\def\QTP#1{}}{}
\@ifundefined{Qlb}{}{}
\@ifundefined{Qlt}{}{}
\long\def\QQA#1#2{}%
\def\QTR#1#2{{\csname#1\endcsname #2}}
\def\EXPAND#1[#2]#3{}%
\def\NOEXPAND#1[#2]#3{}%
\def\LaTeXparent#1{}%
\def\ChildStyles#1{}%
\def\ChildDefaults#1{}%
\def\QTagDef#1#2#3{}%
\def\QQfnmark#1{\footnotemark}

\@ifundefined{INDEX}{\def\INDEX#1#2{}{}}{}%
\@ifundefined{SUBINDEX}{\def\SUBINDEX#1#2#3{}{}{}}{}%
\def\initial#1{\bigbreak{\raggedright\large\bf #1}\kern 2\p@
   \penalty3000}%
\@ifundefined{ZZZ}{}{\makeatletter\input gnuindex.sty\makeatother\makeindex\makeatletter}%
\@ifundefined{abstract}{%
 \def\abstract{%
  \if@twocolumn
   \section*{Abstract (Not appropriate in this style!)}%
   \else \small 
   \begin{center}{\bf Abstract\vspace{-.5em}\vspace{\z@}}\end{center}%
   \quotation 
   \fi
  }%
 }{%
 }%
\@ifundefined{endabstract}{\def\endabstract
  {\if@twocolumn\else\endquotation\fi}}{}%
\@ifundefined{maketitle}{\def\maketitle#1{}}{}%
\@ifundefined{affiliation}{\def\affiliation#1{}}{}%
\@ifundefined{proof}{}{}%
\@ifundefined{endproof}{}{}%
\@ifundefined{newfield}{\def\newfield#1#2{}}{}%
\@ifundefined{chapter}{\def\chapter#1{\par(Chapter head:)#1\par }%
 \newcount\c@chapter}{}%
\@ifundefined{part}{\def\part#1{\par(Part head:)#1\par }}{}%
\@ifundefined{section}{\def\section#1{\par(Section head:)#1\par }}{}%
\@ifundefined{subsection}{\def\subsection#1%
 {\par(Subsection head:)#1\par }}{}%
\@ifundefined{subsubsection}{\def\subsubsection#1%
 {\par(Subsubsection head:)#1\par }}{}%
\@ifundefined{paragraph}{\def\paragraph#1%
 {\par(Subsubsubsection head:)#1\par }}{}%
\@ifundefined{subparagraph}{\def\subparagraph#1%
 {\par(Subsubsubsubsection head:)#1\par }}{}%
\@ifundefined{therefore}{}{}%
\@ifundefined{backepsilon}{}{}%
\@ifundefined{yen}{}{}%
\@ifundefined{registered}{%
   \def\registered{\relax\ifmmode{}\r@gistered
                    \else$\m@th\r@gistered$\fi}%
 \def\r@gistered{^{\ooalign
  {\hfil\raise.07ex\hbox{$\scriptstyle\rm\text{R}$}\hfil\crcr
  \mathhexbox20D}}}}{}%
\@ifundefined{Eth}{}{}%
\@ifundefined{eth}{}{}%
\@ifundefined{Thorn}{}{}%
\@ifundefined{thorn}{}{}%
\@ifundefined{degree}{}{}%
\newdimen\theight
\def\Column{%
 \vadjust{\setbox\z@=\hbox{\scriptsize\quad\quad tcol}%
  \theight=\ht\z@\advance\theight by \dp\z@\advance\theight by \lineskip
  \kern -\theight \vbox to \theight{%
   \rightline{\rlap{\box\z@}}%
   \vss
   }%
  }%
 }%
\def\qed{%
 \ifhmode\unskip\nobreak\fi\ifmmode\ifinner\else\hskip5\p@\fi\fi
 \hbox{\hskip5\p@\vrule width4\p@ height6\p@ depth1.5\p@\hskip\p@}%
 }%
\def\miss{\hbox{\vrule height2\p@ width 2\p@ depth\z@}}%
\def\tcol#1{{\baselineskip=6\p@ \vcenter{#1}} \Column}  %
\def\newfmtname{LaTeX2e}
\def\chkcompat{%
   \if@compatibility
   \else
     \usepackage{latexsym}
   \fi
}

\ifx\fmtname\newfmtname
  \DeclareOldFontCommand{\rm}{\normalfont\rmfamily}{\mathrm}
  \DeclareOldFontCommand{\sf}{\normalfont\sffamily}{\mathsf}
  \DeclareOldFontCommand{\tt}{\normalfont\ttfamily}{\mathtt}
  \DeclareOldFontCommand{\bf}{\normalfont\bfseries}{\mathbf}
  \DeclareOldFontCommand{\it}{\normalfont\itshape}{\mathit}
  \DeclareOldFontCommand{\sl}{\normalfont\slshape}{\@nomath\sl}
  \DeclareOldFontCommand{\sc}{\normalfont\scshape}{\@nomath\sc}
  \chkcompat
\fi

\def\alpha{\Greekmath 010B }%
\def\beta{\Greekmath 010C }%
\def\gamma{\Greekmath 010D }%
\def\delta{\Greekmath 010E }%
\def\epsilon{\Greekmath 010F }%
\def\zeta{\Greekmath 0110 }%
\def\eta{\Greekmath 0111 }%
\def\theta{\Greekmath 0112 }%
\def\iota{\Greekmath 0113 }%
\def\kappa{\Greekmath 0114 }%
\def\lambda{\Greekmath 0115 }%
\def\mu{\Greekmath 0116 }%
\def\nu{\Greekmath 0117 }%
\def\xi{\Greekmath 0118 }%
\def\pi{\Greekmath 0119 }%
\def\rho{\Greekmath 011A }%
\def\sigma{\Greekmath 011B }%
\def\tau{\Greekmath 011C }%
\def\upsilon{\Greekmath 011D }%
\def\phi{\Greekmath 011E }%
\def\chi{\Greekmath 011F }%
\def\psi{\Greekmath 0120 }%
\def\omega{\Greekmath 0121 }%
\def\varepsilon{\Greekmath 0122 }%
\def\vartheta{\Greekmath 0123 }%
\def\varpi{\Greekmath 0124 }%
\def\varrho{\Greekmath 0125 }%
\def\varsigma{\Greekmath 0126 }%
\def\varphi{\Greekmath 0127 }%

\def\nabla{\Greekmath 0272 }

\def\Greekmath#1#2#3#4{%
    \if@compatibility
        \ifnum\mathgroup=\symbold
           \mathchoice{\mbox{\boldmath$\displaystyle\mathchar"#1#2#3#4$}}%
                      {\mbox{\boldmath$\textstyle\mathchar"#1#2#3#4$}}%
                      {\mbox{\boldmath$\scriptstyle\mathchar"#1#2#3#4$}}%
                      {\mbox{\boldmath$\scriptscriptstyle\mathchar"#1#2#3#4$}}%
        \else
           \mathchar"#1#2#3#4%
        \fi 
    \else 
        \ifnum\mathgroup=5 
           \mathchoice{\mbox{\boldmath$\displaystyle\mathchar"#1#2#3#4$}}%
                      {\mbox{\boldmath$\textstyle\mathchar"#1#2#3#4$}}%
                      {\mbox{\boldmath$\scriptstyle\mathchar"#1#2#3#4$}}%
                      {\mbox{\boldmath$\scriptscriptstyle\mathchar"#1#2#3#4$}}%
        \else
           \mathchar"#1#2#3#4%
        \fi     	    
	  \fi}

\newif\ifGreekBold  \GreekBoldfalse
\let\SAVEPBF=\pbf
\def\pbf{\GreekBoldtrue\SAVEPBF}%

\@ifundefined{theorem}{}{}
\@ifundefined{lemma}{}{}
\@ifundefined{corollary}{}{}
\@ifundefined{conjecture}{}{}
\@ifundefined{proposition}{}{}
\@ifundefined{axiom}{}{}
\@ifundefined{remark}{}{}
\@ifundefined{example}{}{}
\@ifundefined{exercise}{}{}
\@ifundefined{definition}{}{}

\@ifundefined{mathletters}{%
  \newcounter{equationnumber}  
  \def\mathletters{%
     \addtocounter{equation}{1}
     \edef\@currentlabel{\theequation}%
     \setcounter{equationnumber}{\c@equation}
     \setcounter{equation}{0}%
     \edef\theequation{\@currentlabel\noexpand\alph{equation}}%
  }
  
}{}

\@ifundefined{BibTeX}{%
    \def\BibTeX{{\rm B\kern-.05em{\sc i\kern-.025em b}\kern-.08em
                 T\kern-.1667em\lower.7ex\hbox{E}\kern-.125emX}}}{}%
\@ifundefined{AmS}%
    {\def\AmS{{\protect\usefont{OMS}{cmsy}{m}{n}%
                A\kern-.1667em\lower.5ex\hbox{M}\kern-.125emS}}}{}%
\@ifundefined{AmSTeX}{}{}%
\ifx\ds@amstex\relax
\makeatother\endinput
\else
   \@ifpackageloaded{amstex}%
      {\message{amstex already loaded}\makeatother\endinput}
      {}
\fi
\let\DOTSI\relax
\def\RIfM@{\relax\ifmmode}%
\def\FN@{\futurelet\next}%
\newcount\intno@
\def\iint{\DOTSI\intno@\tw@\FN@\ints@}%
\def\iiint{\DOTSI\intno@\thr@@\FN@\ints@}%
\def\iiiint{\DOTSI\intno@4 \FN@\ints@}%
\def\idotsint{\DOTSI\intno@\z@\FN@\ints@}%
\def\ints@{\findlimits@\ints@@}%
\newif\iflimtoken@
\newif\iflimits@
\def\findlimits@{\limtoken@true\ifx\next\limits\limits@true
 \else\ifx\next\nolimits\limits@false\else
 \limtoken@false\ifx\ilimits@\nolimits\limits@false\else
 \ifinner\limits@false\else\limits@true\fi\fi\fi\fi}%
\def\multint@{\int\ifnum\intno@=\z@\intdots@                          
 \else\intkern@\fi                                                    
 \ifnum\intno@>\tw@\int\intkern@\fi                                   
 \ifnum\intno@>\thr@@\int\intkern@\fi                                 
 \int}
\def\multintlimits@{\intop\ifnum\intno@=\z@\intdots@\else\intkern@\fi
 \ifnum\intno@>\tw@\intop\intkern@\fi
 \ifnum\intno@>\thr@@\intop\intkern@\fi\intop}%
\def\intic@{%
    \mathchoice{\hskip.5em}{\hskip.4em}{\hskip.4em}{\hskip.4em}}%
\def\negintic@{\mathchoice
 {\hskip-.5em}{\hskip-.4em}{\hskip-.4em}{\hskip-.4em}}%
\def\ints@@{\iflimtoken@                                              
 \def\ints@@@{\iflimits@\negintic@
   \mathop{\intic@\multintlimits@}\limits                             
  \else\multint@\nolimits\fi                                          
  \eat@}
 \else                                                                
 \def\ints@@@{\iflimits@\negintic@
  \mathop{\intic@\multintlimits@}\limits\else
  \multint@\nolimits\fi}\fi\ints@@@}%
\def\intkern@{\mathchoice{\!\!\!}{\!\!}{\!\!}{\!\!}}%
\def\plaincdots@{\mathinner{\cdotp\cdotp\cdotp}}%
\def\intdots@{\mathchoice{\plaincdots@}%
 {{\cdotp}\mkern1.5mu{\cdotp}\mkern1.5mu{\cdotp}}%
 {{\cdotp}\mkern1mu{\cdotp}\mkern1mu{\cdotp}}%
 {{\cdotp}\mkern1mu{\cdotp}\mkern1mu{\cdotp}}}%
\def\RIfM@{\relax\protect\ifmmode}
\def\text{\RIfM@\expandafter\text@\else\expandafter\mbox\fi}
\let\nfss@text\text
\def\text@#1{\mathchoice
   {\textdef@\displaystyle\f@size{#1}}%
   {\textdef@\textstyle\tf@size{\firstchoice@false #1}}%
   {\textdef@\textstyle\sf@size{\firstchoice@false #1}}%
   {\textdef@\textstyle \ssf@size{\firstchoice@false #1}}%
   \glb@settings}

\def\textdef@#1#2#3{\hbox{{%
                    \everymath{#1}%
                    \let\f@size#2\selectfont
                    #3}}}
\newif\iffirstchoice@
\firstchoice@true
\def\Let@{\relax\iffalse{\fi\let\\=\cr\iffalse}\fi}%
\def\vspace@{\def\vspace##1{\crcr\noalign{\vskip##1\relax}}}%
\def\multilimits@{\bgroup\vspace@\Let@
 \baselineskip\fontdimen10 \scriptfont\tw@
 \advance\baselineskip\fontdimen12 \scriptfont\tw@
 \lineskip\thr@@\fontdimen8 \scriptfont\thr@@
 \lineskiplimit\lineskip
 \vbox\bgroup\ialign\bgroup\hfil$\m@th\scriptstyle{##}$\hfil\crcr}%
\def\Sb{_\multilimits@}%
\def\endSb{\crcr\egroup\egroup\egroup}%
\def\Sp{^\multilimits@}%

\newdimen\ex@
\def\rightarrowfill@#1{$#1\m@th\mathord-\mkern-6mu\cleaders
 \hbox{$#1\mkern-2mu\mathord-\mkern-2mu$}\hfill
 \mkern-6mu\mathord\rightarrow$}%
\def\leftarrowfill@#1{$#1\m@th\mathord\leftarrow\mkern-6mu\cleaders
 \hbox{$#1\mkern-2mu\mathord-\mkern-2mu$}\hfill\mkern-6mu\mathord-$}%
\def\leftrightarrowfill@#1{$#1\m@th\mathord\leftarrow
\mkern-6mu\cleaders
 \hbox{$#1\mkern-2mu\mathord-\mkern-2mu$}\hfill
 \mkern-6mu\mathord\rightarrow$}%
\def\overrightarrow{\mathpalette\overrightarrow@}%
\def\overrightarrow@#1#2{\vbox{\ialign{##\crcr\rightarrowfill@#1\crcr
 \noalign{\kern-\ex@\nointerlineskip}$\m@th\hfil#1#2\hfil$\crcr}}}%

\def\overleftarrow{\mathpalette\overleftarrow@}%
\def\overleftarrow@#1#2{\vbox{\ialign{##\crcr\leftarrowfill@#1\crcr
 \noalign{\kern-\ex@\nointerlineskip}$\m@th\hfil#1#2\hfil$\crcr}}}%
\def\overleftrightarrow{\mathpalette\overleftrightarrow@}%
\def\overleftrightarrow@#1#2{\vbox{\ialign{##\crcr
   \leftrightarrowfill@#1\crcr
 \noalign{\kern-\ex@\nointerlineskip}$\m@th\hfil#1#2\hfil$\crcr}}}%
\def\underrightarrow{\mathpalette\underrightarrow@}%
\def\underrightarrow@#1#2{\vtop{\ialign{##\crcr$\m@th\hfil#1#2\hfil
  $\crcr\noalign{\nointerlineskip}\rightarrowfill@#1\crcr}}}%

\def\underleftarrow{\mathpalette\underleftarrow@}%
\def\underleftarrow@#1#2{\vtop{\ialign{##\crcr$\m@th\hfil#1#2\hfil
  $\crcr\noalign{\nointerlineskip}\leftarrowfill@#1\crcr}}}%
\def\underleftrightarrow{\mathpalette\underleftrightarrow@}%
\def\underleftrightarrow@#1#2{\vtop{\ialign{##\crcr$\m@th
  \hfil#1#2\hfil$\crcr
 \noalign{\nointerlineskip}\leftrightarrowfill@#1\crcr}}}%

\def\qopnamewl@#1{\mathop{\operator@font#1}\nlimits@}
\let\nlimits@\displaylimits
\def\setboxz@h{\setbox\z@\hbox}

\def\varlim@#1#2{\mathop{\vtop{\ialign{##\crcr
 \hfil$#1\m@th\operator@font lim$\hfil\crcr
 \noalign{\nointerlineskip}#2#1\crcr
 \noalign{\nointerlineskip\kern-\ex@}\crcr}}}}

 \def\rightarrowfill@#1{\m@th\setboxz@h{$#1-$}\ht\z@\z@
  $#1\copy\z@\mkern-6mu\cleaders
  \hbox{$#1\mkern-2mu\box\z@\mkern-2mu$}\hfill
  \mkern-6mu\mathord\rightarrow$}
\def\leftarrowfill@#1{\m@th\setboxz@h{$#1-$}\ht\z@\z@
  $#1\mathord\leftarrow\mkern-6mu\cleaders
  \hbox{$#1\mkern-2mu\copy\z@\mkern-2mu$}\hfill
  \mkern-6mu\box\z@$}

\def\projlim{\qopnamewl@{proj\,lim}}
\def\injlim{\qopnamewl@{inj\,lim}}
\def\varinjlim{\mathpalette\varlim@\rightarrowfill@}
\def\varprojlim{\mathpalette\varlim@\leftarrowfill@}
\def\varliminf{\mathpalette\varliminf@{}}
\def\varliminf@#1{\mathop{\underline{\vrule\@depth.2\ex@\@width\z@
   \hbox{$#1\m@th\operator@font lim$}}}}
\def\varlimsup{\mathpalette\varlimsup@{}}
\def\varlimsup@#1{\mathop{\overline
  {\hbox{$#1\m@th\operator@font lim$}}}}

\def\stackunder#1#2{\mathrel{\mathop{#2}\limits_{#1}}}%
\begingroup \catcode `|=0 \catcode `[= 1
\catcode`]=2 \catcode `\{=12 \catcode `\}=12
\catcode`\\=12 
|gdef|@alignverbatim#1\end{align}[#1|end[align]]
|gdef|@salignverbatim#1\end{align*}[#1|end[align*]]

|gdef|@alignatverbatim#1\end{alignat}[#1|end[alignat]]
|gdef|@salignatverbatim#1\end{alignat*}[#1|end[alignat*]]

|gdef|@xalignatverbatim#1\end{xalignat}[#1|end[xalignat]]
|gdef|@sxalignatverbatim#1\end{xalignat*}[#1|end[xalignat*]]

|gdef|@gatherverbatim#1\end{gather}[#1|end[gather]]
|gdef|@sgatherverbatim#1\end{gather*}[#1|end[gather*]]

|gdef|@gatherverbatim#1\end{gather}[#1|end[gather]]
|gdef|@sgatherverbatim#1\end{gather*}[#1|end[gather*]]

|gdef|@multilineverbatim#1\end{multiline}[#1|end[multiline]]
|gdef|@smultilineverbatim#1\end{multiline*}[#1|end[multiline*]]

|gdef|@arraxverbatim#1\end{arrax}[#1|end[arrax]]
|gdef|@sarraxverbatim#1\end{arrax*}[#1|end[arrax*]]

|gdef|@tabulaxverbatim#1\end{tabulax}[#1|end[tabulax]]
|gdef|@stabulaxverbatim#1\end{tabulax*}[#1|end[tabulax*]]

|endgroup

\def\align{\@verbatim \frenchspacing\@vobeyspaces \@alignverbatim
You are using the "align" environment in a style in which it is not defined.}

\@namedef{align*}{\@verbatim\@salignverbatim
You are using the "align*" environment in a style in which it is not defined.}
\expandafter\let\csname endalign*\endcsname =\endtrivlist

\def\alignat{\@verbatim \frenchspacing\@vobeyspaces \@alignatverbatim
You are using the "alignat" environment in a style in which it is not defined.}

\@namedef{alignat*}{\@verbatim\@salignatverbatim
You are using the "alignat*" environment in a style in which it is not defined.}
\expandafter\let\csname endalignat*\endcsname =\endtrivlist

\def\xalignat{\@verbatim \frenchspacing\@vobeyspaces \@xalignatverbatim
You are using the "xalignat" environment in a style in which it is not defined.}

\@namedef{xalignat*}{\@verbatim\@sxalignatverbatim
You are using the "xalignat*" environment in a style in which it is not defined.}
\expandafter\let\csname endxalignat*\endcsname =\endtrivlist

\def\gather{\@verbatim \frenchspacing\@vobeyspaces \@gatherverbatim
You are using the "gather" environment in a style in which it is not defined.}

\@namedef{gather*}{\@verbatim\@sgatherverbatim
You are using the "gather*" environment in a style in which it is not defined.}
\expandafter\let\csname endgather*\endcsname =\endtrivlist

\def\multiline{\@verbatim \frenchspacing\@vobeyspaces \@multilineverbatim
You are using the "multiline" environment in a style in which it is not defined.}

\@namedef{multiline*}{\@verbatim\@smultilineverbatim
You are using the "multiline*" environment in a style in which it is not defined.}
\expandafter\let\csname endmultiline*\endcsname =\endtrivlist

\def\arrax{\@verbatim \frenchspacing\@vobeyspaces \@arraxverbatim
You are using a type of "array" construct that is only allowed in AmS-LaTeX.}

\def\tabulax{\@verbatim \frenchspacing\@vobeyspaces \@tabulaxverbatim
You are using a type of "tabular" construct that is only allowed in AmS-LaTeX.}

\@namedef{arrax*}{\@verbatim\@sarraxverbatim
You are using a type of "array*" construct that is only allowed in AmS-LaTeX.}
\expandafter\let\csname endarrax*\endcsname =\endtrivlist

\@namedef{tabulax*}{\@verbatim\@stabulaxverbatim
You are using a type of "tabular*" construct that is only allowed in AmS-LaTeX.}
\expandafter\let\csname endtabulax*\endcsname =\endtrivlist

\def\@@eqncr{\let\@tempa\relax
    \ifcase\@eqcnt \def\@tempa{& & &}\or \def\@tempa{& &}%
      \else \def\@tempa{&}\fi
     \@tempa
     \if@eqnsw
        \iftag@
           \@taggnum
        \else
           \@eqnnum\stepcounter{equation}%
        \fi
     \fi
     \global\tag@false
     \global\@eqnswtrue
     \global\@eqcnt\z@\cr}

 \def\endequation{%
     \ifmmode\ifinner 
      \iftag@
        \addtocounter{equation}{-1} 
        $\hfil
           \displaywidth\linewidth\@taggnum\egroup \endtrivlist
        \global\tag@false
        \global\@ignoretrue   
      \else
        $\hfil
           \displaywidth\linewidth\@eqnnum\egroup \endtrivlist
        \global\tag@false
        \global\@ignoretrue 
      \fi
     \else   
      \iftag@
        \addtocounter{equation}{-1} 
        \eqno \hbox{\@taggnum}
        \global\tag@false%
        $$\global\@ignoretrue
      \else
        \eqno \hbox{\@eqnnum}
        $$\global\@ignoretrue
      \fi
     \fi\fi
 } 

 \newif\iftag@ \tag@false
 
 \def\tag{\@ifnextchar*{\@tagstar}{\@tag}}
 \def\@tag#1{%
     \global\tag@true
     \global\def\@taggnum{(#1)}}
 \def\@tagstar*#1{%
     \global\tag@true
     \global\def\@taggnum{#1}%
}

\makeatother

\begin{document}

\title{\textbf{Linking observables in perturbed topological
field theories }}
\author{\textbf{V. E. R. Lemes, S. P. Sorella, A. Tanzini, O. S. Ventura, } \and 
\textbf{L.C.Q.Vilar, } \\
UERJ, Universidade do Estado do Rio de Janeiro\\
Departamento de F\'\i sica Te\'orica\\
Instituto de F\'{\i}sica\\
Rua S\~ao Francisco Xavier, 524\\
20550-013, Maracan\~{a}, Rio de Janeiro, Brazil \and \textbf{UERJ-DFT/09/99}$%
\,\,\,\,\,\,\,\,\,\,\,\,\,\,\,\,\,\,$\textbf{PACS: 11.10.Gh}}
\maketitle

\begin{abstract}
The topological antisymmetric tensor field theory in $n$-dimensions 
is perturbed by the introduction of local metric dependent 
interaction terms in the curvatures.
The correlator describing the linking number between two surfaces in $n$%
-dimensions is shown to be not affected by the quantum corrections.

\setcounter{page}{0}\thispagestyle{empty}
\end{abstract}

\vfill\newpage\ \makeatother
\renewcommand{\theequation}{\thesection.\arabic{equation}}

\section{Introduction}

Since their introduction \cite{schw, witt}, topological field theories have
been object of continuous and renewed  investigations and have led to many
interesting applications \cite{rep}. Their original motivation was to
provide a natural framework for a field theory characterization and
computation of several topological invariants. For instance, the
Chern-Simons and the BF theories in three dimensions play a relevant role in
knots theory, their correlation functions being respectively related to the
Jones \cite{witt} and the Alexander-Conway \cite{cotta} polynomials. In
particular, a generalization of this framework to higher dimensions allows
to describe the linking number of surfaces in terms of antisymmetric tensor
fields, whose action reads \cite{hor,bt} 
\begin{equation}
\mathcal{S}_{top}=\int_MBdC\;,  \label{s-top}
\end{equation}

\noindent where $B$ and $C$ are forms of degree $p$ 
and $n-p-1$ respectively. Indeed, it is possible to show \cite{hor, bt} 
that the correlation function 

\begin{equation}
\left\langle \int_{\Sigma _p}B\int_{\Sigma _{n-p-1}^{\prime }}C\right\rangle
_{\mathcal{S}_{top}}=Link(\Sigma ,\Sigma ^{\prime })\;  \label{corr-top}
\end{equation}

\noindent gives the linking number between two smooth, closed nonintersecting 
surfaces $\Sigma _p$ and $\Sigma _{n-p-1}^{^{\prime }}$. \noindent It is worth
underlining also that, more recently, much attention has been given to the
fact that topological terms of the kind of (\ref{s-top}) appear frequently
as parts of more general effective actions describing the low-energy
dynamics of several field theory models. For instance, the effective action
corresponding to the bosonization {\cite{hall}} of relativistic
three-dimensional massive fermionic systems at $T=0$ can be written as the
sum of a Chern-Simons term $\int BdB$ and of an infinite series of higher
order terms in the curvature $dB$ and its derivatives. The same procedure
can be applied when fermions are coupled to a dynamical vector field $C$. In
this case the resulting bosonized action contains a mixed term of the type
of (\ref{s-top}). These effective actions are well suited to describe
several three-dimensional phenomena such as 
the Fermi-Bose transmutation \cite{pol}, 
the quantum Hall effect \cite{hall}, the incompressible
chiral liquids \cite{sod,zee}.

\noindent Further interesting examples are provided by the low energy
effective supergravities studied in the context of the $AdS/CFT$
correspondence \cite{ads}. Indeed, as discussed by \cite{ads-witt}, the dual
supergravity action which describes the $N=4\;$SYM theory with electric and
magnetic fluxes is

\begin{equation}
\mathcal{S}_{sugra}=\int_XB^{RR}dB^{NS}\;+\;\mathcal{S}(dB)\;,  \label{s-g}
\end{equation}

\noindent where {\ }$B^{RR},\;B^{NS}\;$are the Ramond-Ramond and
Neveu-Schwarz--Neveu-Schwarz two forms, transforming as a doublet under the $%
SL(2,\Bbb{Z})$ S-duality symmetry, $X$ is a negatively curved five-dimensional 
Einstein manifold, and $\mathcal{S}(dB)$ collects all the higher
order terms in the curvatures $dB^{RR},\;dB^{NS}$.

\noindent In all these models, the topological terms appear together with
higher derivatives terms in the generalized curvatures, which depend on the
metric. Therefore we are naturally led to analyse the possible contributions
coming from these terms, as recently done for the three-dimensional abelian
Chern-Simons theory {\cite{pert}. In this case it has been possible to prove
that }the correlator

\begin{equation}
\left\langle \int_{\gamma _1}A\int_{\gamma _2}A\right\rangle =Link(\gamma
_1,\gamma _2)\;,  \label{c-s-l}
\end{equation}

\noindent describing the linking number between two smooth closed 
nonintersecting curves $\gamma _1,\gamma _2$, is not affected 
by the presence of higher derivative interaction terms of 
the type $\int F^n$.

\noindent The aim of the present work is that of generalizing this result
to higher dimensions. In particular we shall see that, in analogy with the
three-dimensional Chern-Simons, the correlator

\begin{equation}
\left\langle \int_{\Sigma _p}B\int_{\Sigma _{n-p-1}^{\prime }}C\right\rangle
_{\left( \mathcal{S}_{top}+\mathcal{S}_{int}\right) }=Link(\Sigma ,\Sigma
^{\prime })\;,  \label{1}
\end{equation}

\noindent is not affected by the introduction of local perturbation terms 
in the curvatures of the type 
\begin{equation}
\mathcal{S}_{int}=\int \;(dB)^k(dC)^q\;,  \label{int}
\end{equation}

\noindent and gives as final result the linking number of the 
two surfaces $\Sigma,\Sigma ^{\prime }$. This result means that 
the correlator (\ref{1}) is stable against the introduction of 
perturbing local terms in the curvatures,
regardless of their power-counting nonrenormalizability.

\noindent It is worth underlining that we will limit here ourselves only 
to effective actions which are abelian, and we shall consider only terms 
dependent on the curvatures $H_B=dB,\;H_C=dC\;$ which can be treated as 
true perturbations.
Namely, we shall avoid in the effective action $\left( \mathcal{S}_{top}+%
\mathcal{S}_{int}\right) $ the inclusion of a term of the Maxwell type 
\begin{equation}
\mathcal{S}_{Max}=\frac 1{m_B}\int_MH_B*H_B+\frac 1{m_C}\int_MH_C*H_C\,\;,
\label{max}
\end{equation}

\noindent where $m_B,$ $m_C$ are mass parameters. The presence of 
this term would completely modify the original properties of 
the model. In fact, being expression (\ref{max}) quadratic in the 
gauge fields, it cannot be considered as a true perturbation term, as it 
will be responsible for the presence of massive excitations in the spectrum 
of the theory. Rather, the presence of the Maxwell term in the effective 
action $\left( \mathcal{S}_{top}+\mathcal{S}_{int}\right)$ will give rise 
to the existence of two distinct regimes corresponding to the long and short 
distance behaviors, respectively. For distances larger than the inverse 
of the mass parameters $m_B,$ $m_C\;$ (\textit{i.e.}, the low energy regime ), 
the topological term
will prevail,\textit{\ }while the Maxwell term will become the relevant one
at short distances (\textit{i.e.,} the high-energy regime). In general, the
computation of the correlator (\ref{corr-top}) for arbitrary surfaces in the
presence of the Maxwell term is a very difficult task, mainly due to the
metric dependence of the propagator. Indeed, in the abelian
three-dimensional Maxwell-Chern-Simons case no closed analytic expression
for the correlator (\ref{c-s-l}) has been found so far, even in the simpler
case in which the curves $\gamma _1,\gamma _2$ are two circles. The
inclusion of the Maxwell term is beyond the aim of the present paper, being
presently under investigation. We would like to remark that, in the low
energy regime, a useful local gauge invariant field redefinition has been
recently applied {\cite{slk}} for the Maxwell-Chern-Simons theory to compute
the corrections to the correlator (\ref{c-s-l}) for two coinciding curves $%
\gamma _1\equiv \gamma _2$ as power series in the inverse of the mass. This
field redefinition can be also generalized to the present case and is
expected to be suitable for the characterization of the Maxwell
contributions in the low energy regime {\cite{w-p}}.

\section{Perturbative expansion and Feynman diagrams}

In order to analyse the aforementioned nonrenormalization properties of the
linking number, let us proceed by computing the correlator 
\begin{equation}
\left\langle \int_{\Sigma _p}B\int_{\Sigma _{n-p-1}^{\prime }}C\right\rangle
_{\mathcal{S}eff}\;,  \label{corr}
\end{equation}

\noindent in the case in which the total action $\mathcal{S}$ is chosen to be
\begin{eqnarray}
\mathcal{S}_{eff} &=&\mathcal{S}_{top}+\mathcal{S}_{int}\;,  \nonumber  
\label{6}
\\
\mathcal{S}_{int} &=&\tau \int_M:(H_B*H_B)*(H_C*H_C):\;  \label{action}
\end{eqnarray}

\noindent where $\tau $ is an arbitrary parameter with negative mass dimension, and $%
H_B=dB$, $H_C=dC$ are respectively the curvatures of the gauge fields $B$
and $C$. Notice that the normal ordering prescription has been adopted for
the quartic interaction, allowing us to rule out tadpoles diagrams\footnote{%
We remind the reader that, in the present abelian case, the normal-ordering
prescription is compatible with the requirement of gauge invariance. This
follows from the observation that the positive and negative-frequency parts
of the field-strengths related to the gauge fields $C$ and $B$ are each
gauge invariant.}.

\noindent The action $\mathcal{S}$ is invariant under the gauge
transformations
\begin{eqnarray}
\delta B &=&d\omega _{p-1\;,}  \nonumber \\
\delta C &=&d\eta _{n-p-2}\;.  \label{g-t}
\end{eqnarray}
Remark that the transformations $(\ref{g-t})$ are reducible and therefore
the quantization procedure will require the introduction of ghosts for
ghosts. In order to gauge fix the action $(\ref{action})$ we choose the
linear Landau tansverse condition

\begin{equation}
d*B=d*C=0\;,  \label{g-f}
\end{equation}
which implies that the whole set of ghosts for ghosts completely decouples,
due to the abelian character of the theory. Thus, for the propagator we get

\begin{equation}
\left\langle B_{\mu _1...\mu _p}(x)C_{\mu _{p+1}...\mu _{n-1}}(y)\right\rangle _{BC}=
\varepsilon _{\mu _1......\mu _{n-1}\mu
_n}\partial _x^{\mu _n}G(x-y)\;,  \label{prop}
\end{equation}
where $G(x-y)$ is the Green function of the laplacian 
\[
\ast d*dG(x-y)=\delta (x-y)\;.
\]
Since the two surfaces $\Sigma ,\;\Sigma ^{\prime }$ must be homologically
trivial \cite{hor}

\[
\Sigma _p=\partial \mathcal{\beta }_{p+1},\Sigma _{n-p-1}^{\prime }=\partial 
\mathcal{\beta }_{n-p}\;,
\]
we can suppose that they are contained in a region of $M$ diffeomorphic to
the flat euclidean space\footnote{%
Actually, there are some counterexamples of surfaces which are homologically
trivial but cannot be contained in a region diffeomorphic to $R ^n$ \cite
{hor}, but we do not consider these particular cases here.} $R ^n$. In
this case, the Green function is given by

\[
G(x-y)=\frac{\Gamma \left( n/2\right) }{\left[ \left( 2n-4\right) \pi
^{n/2}p!\left( n-p-1\right) !\right] }\frac{^1}{\left| x-y\right| ^{n-2}}\,, 
\]
where $\left| x-y\right| $ denotes the euclidean distance between $x$ and $y$%
.
\noindent Let us now give the elementary Wick contractions which shall be
needed for the evaluation of the Feynman diagrams. From $(\ref{prop})$ we
obtain
\begin{eqnarray}
\left\langle B_{\mu _1...\mu _p}(x)\widetilde{H}_C^{\nu _1...\nu
_p}(y)\right\rangle  &=&\left( -1\right) ^{n+p}(p+1)\left( \delta _{\left[
\mu _1...\mu _p\right] }^{\nu _1...\nu _p}\delta (x-y)+\right.   \nonumber \\
&&\left. +\,\,\,\,\delta _{[\mu _1,..,\mu _{i-1},\mu _{i+1},..,\mu _p}^{\nu
_1,..,\nu _{i-1},\nu _{i+1},..,\nu _p}\partial _{\mu _i]}^x\partial _y^{\nu
_i}G(x-y)\right), \nonumber \\  \label{bhcprop}
\end{eqnarray}
\begin{eqnarray}
\left\langle C_{\mu _1...\mu _{n-p-1}}(x)\widetilde{H}_B^{\nu _1...\nu
_{n-p-1}}(y)\right\rangle  &=&\frac{p!}{(n-p-1)!}p\left( \delta _{\left[ \mu
_1...\mu _{n-p-1}\right] }^{\nu _1...\nu _{n-p-1}}\delta (x-y)+\right. 
  \nonumber \\&&\left. +\,\,\,\,\delta _{[\mu _1,..,\mu _{i-1},\mu _{i+1},..,\mu
_{n-p-1}}^{\nu _1,..,\nu _{i-1},\nu _{i+1},..,\nu _{n-p-1}}\partial _{\mu
_i]}^x\partial _y^{\nu _i}G(x-y)\right) ,  \nonumber \\ \label{chbprop}
\end{eqnarray}
where the index $i$ runs from one to $p$, and 
\begin{eqnarray}
\left\langle \widetilde{H}_{B\mu _1...\mu _{n-p-1}}(x)\widetilde{H}_{C\nu
_1...\nu _p}(y)\right\rangle &=&\frac{(-1)^{p(n-p)+n-1}p(n-p)^2}{(p-1)!}\times
\nonumber \\&& \times \left( \varepsilon _{\mu _1...\mu _{n-p-1}\nu _1...\nu _p\nu
_{p+1}}^{\,\,\,\,\,\,\,\,\,\,\,\,\,\,\,\,\,\,\,\,\,\,\,\,}\partial _y^{\nu
_{p+1}}\delta \left( x-y\right) \right) .  \nonumber \\ \label{hbhcprop}
\end{eqnarray}

\noindent Notice that the propagator $\left\langle C\widetilde{H}_B\right\rangle $ has
the same structure of $\left\langle B\widetilde{H}_C\right\rangle $ in $(%
\ref{bhcprop})$. In the above equations $\widetilde{H}_C=*H_C$ and $%
\widetilde{H}_B=*H_B$ are respectively the Hodge duals of the curvatures $H_C
$ and $H_B,$ and the symbol $[\mu _1,...,\mu _p]$ means complete
antisymmetrization in the indices.

\noindent Concerning now the perturbative expansion, it is easy to see that
the first Feynman diagram which contributes to the correlation function $(%
\ref{corr})$ is of two-loop order. For the three-dimensional case, in which $%
\Sigma ,$ $\Sigma ^{\prime }$ are two smooth oriented nonintersecting
curves, this diagram can be drawn as follows
\begin{center}
\begin{figure}[h]
\setlength{\unitlength}{1mm} 
\hspace{0.5cm}\scalebox{0.75}{\includegraphics*{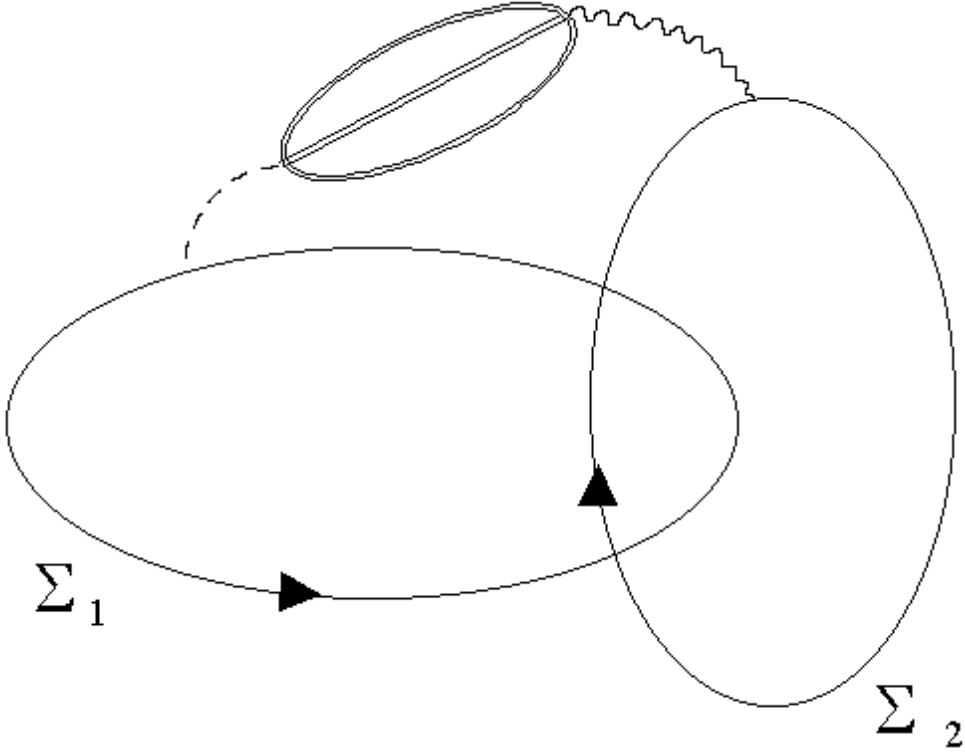}}
\caption{2-loop contribution}
\end{figure}
\end{center}
\noindent In the above figure, the dashed, the wavy and the double lines refer
respectively to the Wick contractions $\left\langle B\widetilde{H}%
_C\right\rangle ,\left\langle C\widetilde{H}_B\right\rangle $ and $%
\left\langle \widetilde{H}_B\widetilde{H}_C\right\rangle $.

\noindent Apart from an irrelevant global symmetry coefficient, the Feynman
integral corresponding to the diagram of $Fig.1$ is given by 
\begin{eqnarray}
I^{\left( 2\right) } &=&\int_\Sigma d\sigma \left( x\right) ^{\mu _1....\mu
_p}\int_{\Sigma ^{\prime }}d\tau (y)^{\nu _1....\nu _{n-p-1}}\int d^nz_1\int
d^nz_2  \label{i2} \nonumber \\
&&\left\langle B_{\mu _1....\mu _p}\left( x\right) \widetilde{H}_{C\beta
_1....\beta _p}\left( z_1\right) \right\rangle \left\langle C_{\nu _1....\nu
_{n-p-1}}(y)\widetilde{H}_{B\gamma _1....\gamma _{n-p-1}}\left( z_2\right)
\right\rangle   \nonumber \\
&&\left\langle \widetilde{H}_C^{\beta _1....\beta _p}\left( z_1\right) 
\widetilde{H}_B^{\gamma _1....\gamma _{n-p-1}}\left( z_2\right)
\right\rangle \left\langle \widetilde{H}_B^{\lambda _1....\lambda
_{n-p-1}}\left( z_1\right) \widetilde{H}_C^{\alpha _1....\alpha _p}\left(
z_2\right) \right\rangle   \nonumber \\
&&\left\langle \widetilde{H}_{B\lambda _1....\lambda _{n-p-1}}\left(
z_1\right) \widetilde{H}_{C\alpha _1....\alpha _p}\left( z_2\right)
\right\rangle \;\,.  
\end{eqnarray}

\noindent In the analysis of the above expression it is worth to observe that the two
fields $B(x)$ and $C(y)$, lying on the hypersurfaces $\Sigma $ and $\Sigma
^{\prime }$, are respectively Wick-contracted with $\widetilde{H}_C$ and $%
\widetilde{H}_B$. From the explicit form of the propagator $\left\langle B%
\widetilde{H}_C\right\rangle $, we see that the term in the second line
of $(\ref{bhcprop})$ yields a total derivative on the closed hypersurface 
$\Sigma $; implying that this term gives a vanishing contribution. The
same result also holds for the contribution coming from the propagator $%
\left\langle C\widetilde{H}_B\right\rangle $ \ of $(\ref{chbprop})$. It
is worth remarking that this argument relies only on the fact that the two
surfaces $\Sigma $ and $\Sigma ^{\prime }$ are closed. Therefore, it applies
to the higher-order contributions as well. As a consequence, all the Wick
contractions entering the Feynman diagrams will lead to a product of delta
functions. In particular, for the 2-loop diagram of $Fig.1$, the integral $(%
\ref{i2})$ reduces to 
\begin{eqnarray}
I^{(2)} &=&\int d\sigma (x)^{\mu _1....\mu _p}\int d\tau (y)^{\nu _1....\nu
_{n-p-1}}\int d^nz_1\int d^nz_2  \label{i22} \nonumber \\
&&\delta _{\left[ \nu _1....\nu _{n-p-1}\right] }^{\gamma _1....\gamma
_{n-p-1}}\delta \left( y-z_2\right) \delta _{\left[ \mu _1....\mu _p\right]
}^{\sigma _1....\sigma _p}\delta \left( x-z_1\right)   \nonumber \\
&&\varepsilon ^{\sigma _1....\sigma _{p+1}\gamma _1....\gamma
_{n-p-1}}\partial _{\sigma _{p+1}}\delta \left( z_1-z_2\right) \varepsilon
^{\alpha _1....\alpha _{p+1}\lambda _1....\lambda _{n-p-1}\,}\partial
_{\alpha _{p+1}}\delta \left( z_1-z_2\right)   \nonumber \\
&&\varepsilon _{\alpha _1....\alpha _p\widehat{\alpha }_{p+1}\lambda
_1....\lambda _{n-p-1}}^{\,}\partial ^{\widehat{\alpha }_{p+1}}\delta \left(
z_1-z_2\right) .  
\end{eqnarray}

\noindent In spite of the presence of products of delta functions with the same
argument, the above expression is easily seen to vanish. We observe first of all 
that it is always possible to take an order of integration over
the delta functions such that we end up with products of $\delta ^n(x-y)$
and not of $\delta ^n\left( 0\right) $. In the present case, this would
amount to integrate out first the two delta functions with arguments $x-z_1$
and $y-z_2$, leading to 

\begin{eqnarray}
I^{(2)} &=&\int d\sigma (x)^{\mu _1....\mu _p}\int d\tau (y)^{\nu _1....\nu
_{n-p-1}}\delta _{\left[ \nu _1....\nu _{n-p-1}\right] }^{\gamma
_1....\gamma _{n-p-1}}\delta _{\left[ \mu _1....\mu _p\right] }^{\sigma
_1....\sigma _p}  \nonumber \\
&&\varepsilon ^{\sigma _1....\sigma _{p+1}\gamma _1....\gamma
_{n-p-1}}\varepsilon ^{\alpha _1....\alpha _{p+1}\lambda _1....\lambda
}\varepsilon _{\alpha _1....\alpha _p\widehat{\alpha }_{p+1}\lambda
_1....\lambda _{n-p-1}}^{\,}  \nonumber \\
&&\partial _{\sigma _{p+1}}\delta \left( x-y\right) \partial _{\alpha
_{p+1}}\delta \left( x-y\right) \partial ^{\widehat{\alpha }_{p+1}}\delta
\left( x-y\right) ,
\end{eqnarray}

\noindent which is zero owing to the fact that the two surfaces $\Sigma $ and $\Sigma
^{\prime }$ never intersect, so that $x-y$ never vanishes. Moreover, as done
in $\cite{pert}$, we can adopt a more rigorous treatment by regularizing the 
$n$-dimensional delta functions with coinciding arguments through the point
splitting procedure already used by Polyakov $\cite{pol}$%
\begin{eqnarray}
\delta _\epsilon (z_1-z_2) &=&\frac 1{\left( 2\pi \epsilon \right)
^{n/2}}e^{-\left| z_1-z_2\right| ^2/2\epsilon }  \nonumber \\
\stackunder{\epsilon \rightarrow 0}{\lim }\delta _\epsilon (z_1-z_2)
&=&\delta (z_1-z_2) \label{reg}  
\end{eqnarray}
\noindent More precisely, whenever a product of $m$ delta functions with coinciding
arguments occurs, it will be understood as 
\[
\left[ \delta \left( z_1-z_2\right) \right] ^m=\left[ \delta _\epsilon
\left( z_1-z_2\right) \right] ^{m-1}\delta \left( z_1-z_2\right) ,
\]
where the limit $\epsilon \rightarrow 0$ is meant to be taken at the end of
all calculations.
\noindent In this regularization scheme we then have 

\begin{eqnarray}
I^{(2)} &=&\stackunder{\epsilon \rightarrow 0}{\lim }\int d\sigma (x)^{\mu
_1....\mu _p}\int d\tau (y)^{\nu _1....\nu _{n-p-1}}\int d^nz_1\int d^nz_2
 \nonumber \\
&&\delta _{\left[ \nu _1....\nu _{n-p-1}\right] }^{\gamma _1....\gamma
_{n-p-1}}\delta (y-z_2)\delta _{\left[ \mu _1....\mu _p\right] }^{\sigma
_1....\sigma _p}\delta \left( x-z_1\right)  \nonumber \\
&&\varepsilon ^{\sigma _1....\sigma _{p+1}\gamma _1....\gamma
_{n-p-1}}\partial _{\sigma _{p+1}}\delta _\epsilon \left( z_1-z_2\right)
\varepsilon ^{\alpha _1....\alpha _{p+1}\lambda _1....\lambda
_{n-p-1}\,}\partial _{\alpha _{p+1}}\delta _\epsilon \left( z_1-z_2\right) 
\nonumber \\
&&\varepsilon _{\alpha _1....\alpha _p\widehat{\alpha }_{p+1}\lambda
_1....\lambda _{n-p-1}}^{\,}\partial ^{\widehat{\alpha }_{p+1}}\delta \left(
z_1-z_2\right) , \label{i22reg} 
\end{eqnarray}

\noindent which is vanishing no matter the order of integration we take. Indeed,
before taking the limit $\epsilon \rightarrow 0,$ we always obtain an
expression containing $\delta \left( x-y\right) $, with $x-y\neq 0$. We can
thus conclude that the two-loop diagram in $Fig.1$ does not contribute to
the correlator $(\ref{corr})$.

\noindent Similar arguments can be used to show that also the higher-order
contributions vanish. Let us consider, for instance, the topologically
distinct diagrams contributing to the 3 and 4-loop order, shown respectively
in $Fig.2$ and $Figs.3,4$.
\begin{center}
\begin{figure}[h]
\setlength{\unitlength}{1mm} 
\hspace{1.0cm}\scalebox{0.65}{\includegraphics*{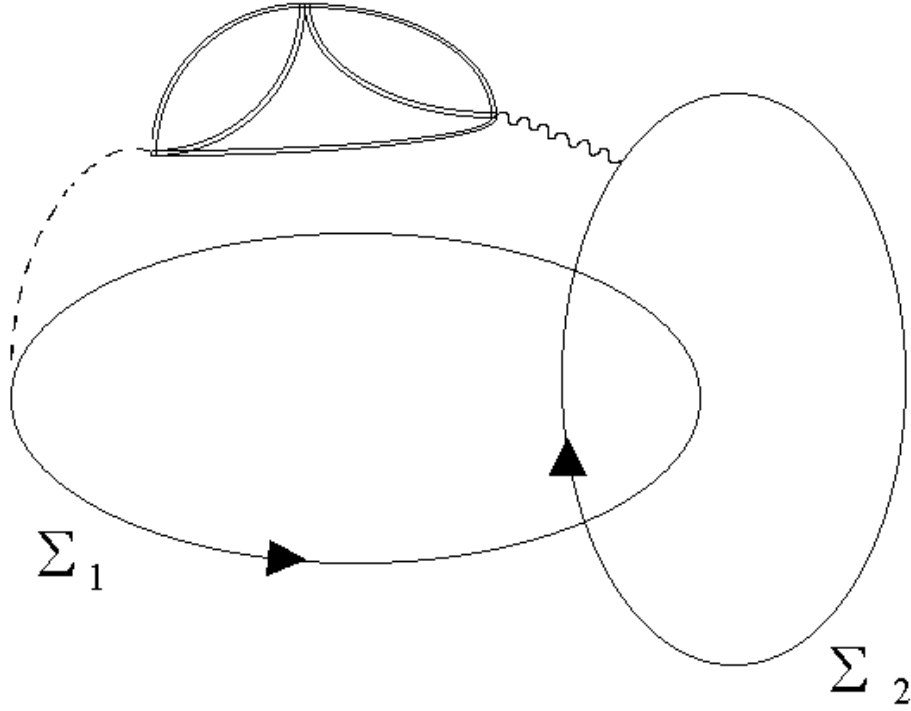}}
\caption{3-loop contribution}
\end{figure}
\end{center}

\begin{center}
\begin{figure}[h]
\setlength{\unitlength}{1mm}
\hspace{-3.2cm}\scalebox{1.1}{\includegraphics*{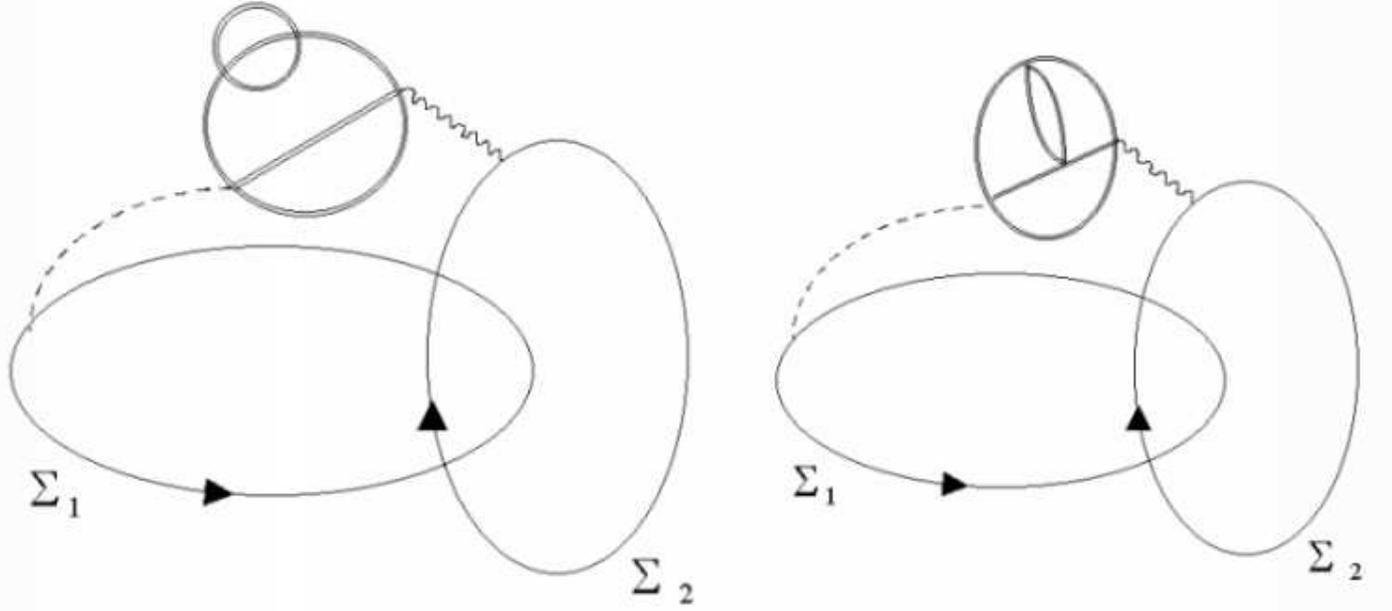}}
\caption{4-loop contributions}
\end{figure}
\end{center}

\noindent Concerning the 3-loop contribution of $Fig.2$, a typical 
contraction is 

\begin{eqnarray}
I^{(3)} &=&\int d\sigma \left( x\right) ^{\mu _1....\mu _p}\int d\tau
(y)^{\nu _1....\nu _{n-p-1}}\int d^nz_1\int d^nz_2\int d^nz_3  \nonumber \\
&&\;\;\;\;\;\delta _{\left[ \nu _1....\nu _{n-p-1}\right] }^{\varphi
_1....\varphi _{n-p-1}}\delta (y-z_2)\delta _{\left[ \mu _1....\mu _p\right]
}^{\beta _1....\beta _p}\delta \left( x-z_1\right) \varepsilon _{\rho
_1...\rho _{p+1}\alpha _1...\alpha
_{n-p-1}}^{\,\,\,\,\,\,\,\,\,\,\,\,\,\,\,\,\,\,\,\,\,\,\,\,}\partial ^{\rho
_{p+1}}\delta \left( z_1-z_3\right)   \nonumber \\
&&\;\;\;\;\;\varepsilon _{\sigma _1...\sigma _{p+1}\alpha _1...\alpha
_{n-p-1}}^{\,\,\,\,\,\,\,\,\,\,\,\,\,\,\,\,\,\,\,\,\,\,\,\,}\partial
^{\sigma _{p+1}}\delta \left( z_1-z_2\right) \varepsilon _{\beta _1....\beta
_{p+1}\theta _1...\theta
_{n-p-1}}^{\,\,\,\,\,\,\,\,\,\,\,\,\,\,\,\,\,\,\,\,\,\,\,\,}\partial ^{\beta
_{p+1}}\delta \left( z_1-z_3\right)   \nonumber \\
&&\;\;\;\;\;\varepsilon ^{\sigma _1...\sigma _p\widehat{\sigma }_{p+1}\theta
_1...\theta _{n-p-1}}\partial _{\widehat{\sigma }_{p+1}}\delta \left(
z_3-z_2\right) \varepsilon _{\,\,\,\,\,\,\,\,\,\,\,\,\,\,\,\,\,\,\varphi
_1....\varphi _{n-p-1}}^{\rho _1...\rho _p\widehat{\rho }_{p+1}\,\,\,\,\,\,%
\,\,\,\,\,\,\,\,\,\,\,\,\,\,\,\,\,\,}\partial _{\widehat{\sigma }%
_{p+1}}\delta \left( z_2-z_3\right) \;. \nonumber \\ \label{13}
\end{eqnarray}
\newline

\noindent Analogously, a typical term of the four-loop contribution of $Fig.3,$ turns
out to be  

\begin{eqnarray}
I^{\left( 4\right) } &=&\int d\sigma \left( x\right) ^{\mu _1....\mu _p}\int
d\tau (y)^{\nu _1....\nu _{n-p-1}}\int d^nz_1\int d^nz_2\int d^nz_3\int
d^nz_4  \nonumber \\
&&_{\left[ \mu _1....\mu _p\right] }^{\sigma _1....\sigma _p}\delta \left(
x-z_1\right) \delta _{\left[ \nu _1....\nu _{n-p-1}\right] }^{\beta
_1....\beta _{n-p-1}}\delta (y-z_2)\varepsilon ^{\theta _1...\theta
_{p+1}\alpha _1...\alpha _{n-p-1}}\partial _{\theta _{p+1}}\delta \left(
z_1-z_2\right)   \nonumber \\
&&\varepsilon _{\theta _1...\theta _p\widehat{\theta }_{p+1}\alpha
_1...\alpha _{n-p-1}}\partial ^{\widehat{\theta }_{p+1}}\delta \left(
z_1-z_2\right) \varepsilon _{\sigma _1...\sigma _{p+1}\varphi _1...\varphi
_{n-p-1}}^{\,\,\,\,\,\,\,\,\,\,\,\,\,\,\,\,\,\,\,\,\,\,\,\,}\partial
^{\sigma _{p+1}}\delta \left( z_1-z_3\right)   \nonumber \\
&&\varepsilon _{\xi _1...\xi _{p+1}\beta _1...\beta
_{n-p-1}}^{\,\,\,\,\,\,\,\,\,\,\,\,\,\,\,\,\,\,\,\,\,\,\,\,}\partial ^{\xi
_{p+1}}\delta \left( z_2-z_4\right) \varepsilon _{\xi _1...\xi _p\widehat{%
\xi }_{p+1}\varphi _1...\varphi
_{n-p-1}}^{\,\,\,\,\,\,\,\,\,\,\,\,\,\,\,\,\,\,\,\,\,\,\,\,}\partial ^{%
\widehat{\xi }_{p+1}}\delta \left( z_3-z_4\right)   \nonumber \\
&&\varepsilon ^{\eta _1...\eta _{p+1}\gamma _1...\gamma _{n-p-1}}\partial
_{\eta _{p+1}}\delta \left( z_3-z_4\right) \varepsilon _{\eta _1...\eta _p%
\widehat{\eta }_{p+1}\gamma _1...\gamma
_{n-p-1}}^{\,\,\,\,\,\,\,\,\,\,\,\,\,\,\,\,\,\,\,\,\,\,\,\,}\partial ^{%
\widehat{\eta }_{p+1}}\delta \left( z_3-z_4\right) \;.  \nonumber \\ \label{14} 
\end{eqnarray}

\noindent Integrating over $z_1$ and $z_2$, and after regularizing the product of
delta functions with coinciding arguments, we obtain
\newline

\begin{eqnarray}
I^{(3)} &=&\int d\sigma \left( x\right) ^{\mu _1....\mu _p}\int d\tau
(y)^{\nu _1....\nu _{n-p-1}}\int d^nz_3 \nonumber \\
&&\;\;\;\;\;\delta _{\left[ \nu _1....\nu _{n-p-1}\right] }^{\varphi
_1....\varphi _{n-p-1}}\delta _{\left[ \mu _1....\mu _p\right] }^{\beta
_1....\beta _p}\varepsilon _{\rho _1...\rho _{p+1}\alpha _1...\alpha
_{n-p-1}}^{\,\,\,\,\,\,\,\,\,\,\,\,\,\,\,\,\,\,\,\,\,\,\,\,}\partial ^{\rho
_{p+1}}\delta \left( x-z_3\right)  \nonumber \\
&&\;\;\;\;\;\varepsilon _{\sigma _1...\sigma _{p+1}\alpha _1...\alpha
_{n-p-1}}^{\,\,\,\,\,\,\,\,\,\,\,\,\,\,\,\,\,\,\,\,\,\,\,\,}\partial
^{\sigma _{p+1}}\delta \left( x-y\right) \varepsilon _{\beta _1....\beta
_{p+1}\theta _1...\theta
_{n-p-1}}^{\,\,\,\,\,\,\,\,\,\,\,\,\,\,\,\,\,\,\,\,\,\,\,\,}\partial ^{\beta
_{p+1}}\delta \left( x-z_3\right)  \nonumber \\
&&\;\;\;\;\;\varepsilon ^{\sigma _1...\sigma _p\widehat{\sigma }_{p+1}\theta
_1...\theta _{n-p-1}}\partial _{\widehat{\sigma }_{p+1}}\delta \left(
z_3-y\right) \varepsilon _{\,\,\,\,\,\,\,\,\,\,\,\,\,\,\,\,\,\,\varphi
_1....\varphi _{n-p-1}}^{\rho _1...\rho _p\widehat{\rho }_{p+1}\,\,\,\,\,\,%
\,\,\,\,\,\,\,\,\,\,\,\,\,\,\,\,\,\,}\partial _{\widehat{\sigma }%
_{p+1}}\delta _\epsilon \left( y-z_3\right) \;,  \nonumber \\ \label{13-reg}
\end{eqnarray}
and 
\begin{eqnarray}
I^{\left( 4\right) } &=&\int d\sigma \left( x\right) ^{\mu _1....\mu _p}\int
d\tau (y)^{\nu _1....\nu _{n-p-1}}\int d^nz_3\int d^nz_4 \nonumber \\
&&_{\left[ \mu _1....\mu _p\right] }^{\sigma _1....\sigma _p}\delta _{\left[
\nu _1....\nu _{n-p-1}\right] }^{\beta _1....\beta _{n-p-1}}\varepsilon
^{\theta _1...\theta _{p+1}\alpha _1...\alpha _{n-p-1}}\partial _{\theta
_{p+1}}\delta \left( x-y\right)  \nonumber \\
&&\varepsilon _{\theta _1...\theta _p\widehat{\theta }_{p+1}\alpha
_1...\alpha _{n-p-1}}\partial ^{\widehat{\theta }_{p+1}}\delta \left(
x-y\right) \varepsilon _{\sigma _1...\sigma _{p+1}\varphi _1...\varphi
_{n-p-1}}^{\,\,\,\,\,\,\,\,\,\,\,\,\,\,\,\,\,\,\,\,\,\,\,\,}\partial
^{\sigma _{p+1}}\delta \left( x-z_3\right)  \nonumber \\
&&\varepsilon _{\xi _1...\xi _{p+1}\beta _1...\beta
_{n-p-1}}^{\,\,\,\,\,\,\,\,\,\,\,\,\,\,\,\,\,\,\,\,\,\,\,\,}\partial ^{\xi
_{p+1}}\delta \left( y-z_4\right) \varepsilon _{\xi _1...\xi _p\widehat{\xi }%
_{p+1}\varphi _1...\varphi
_{n-p-1}}^{\,\,\,\,\,\,\,\,\,\,\,\,\,\,\,\,\,\,\,\,\,\,\,\,}\partial ^{%
\widehat{\xi }_{p+1}}\delta \left( z_3-z_4\right)  \nonumber \\
&&\varepsilon ^{\eta _1...\eta _{p+1}\gamma _1...\gamma _{n-p-1}}\partial
_{\eta _{p+1}}\delta _\epsilon \left( z_3-z_4\right) \varepsilon _{\eta
_1...\eta _p\widehat{\eta }_{p+1}\gamma _1...\gamma
_{n-p-1}}^{\,\,\,\,\,\,\,\,\,\,\,\,\,\,\,\,\,\,\,\,\,\,\,\,}\partial ^{%
\widehat{\eta }_{p+1}}\delta _\epsilon \left( z_3-z_4\right) \;.  
\nonumber \\ \label{14-reg}
\end{eqnarray}
All terms in all possible diagrams may then be seen to be proportional to
the $n$-dimensional delta function $\delta \left( x-y\right) $ or its
derivatives. One may easily convince oneself that this mechanism also
applies to any order in perturbation theory. Since we always have $x\neq y$,
these diagrams all amount to a null correction to the basic diagram, so that
the correlation function $(\ref{corr})$ gives the generalized linking number
of $\Sigma $ and $\Sigma ^{\prime }$

\begin{equation}
\left\langle \int_{\Sigma _p}B\int_{\Sigma _{n-p-1}^{\prime }}C\right\rangle
_{\mathcal{S}_{eff}}=Link\left( \Sigma ,\Sigma ^{\prime }\right) \;.
\label{15}
\end{equation}

\section{Conclusion}

We have been able to prove that the correlator $(\ref{corr})$ is unaffected
by radiative corrections to all orders, provided that $\Sigma $ and $\Sigma
^{\prime }$ are two smooth closed nonintersecting surfaces. Although we have
carried out explicit computations for the interaction term $\int H_B^2H_C^2$%
,\ the same result may be achieved straightforwardly for any local
interaction containing arbitrary powers of the curvatures $\int H_B^pH_C^q$.
This result, which generalizes to arbitrary dimensions that obtained in the
case of two nonintersecting curves $\cite{pert}$, may be interpreted as a
kind of nonrenormalization property of the  linking number in higher
dimensions. 

\vspace{2cm}

{\Large \textbf{Acknowledgements}}

We are indebted to the Departamento de Campos e Part{\'\i}culas (DCP)\ of
the Centro Brasileiro de Pesquisas F{\'\i}sicas (CBPF)\ for kind
hospitality. The Conselho\ Nacional de Pesquisa e Desenvolvimento
(CNPq/Brazil), the Funda\c {c}\~{a}o de Amparo \`{a} Pesquisa do Estado do
Rio de Janeiro (Faperj) and the SR2-UERJ are gratefully acknowledged for
financial support.

\end{document}